\documentclass[aps,pra,twocolumn,showpacs,amsmath,amssymb,preprintnumbers,superscriptaddress,10pt]{revtex4-2}

\usepackage{graphicx}
\usepackage{graphicx}
\usepackage{SIunits}
\usepackage{hyphenat}
\usepackage{braket}
\usepackage{appendix}
\usepackage[bookmarks=false]{hyperref}
\usepackage{color}
\usepackage[capitalise]{cleveref}
\usepackage{amsmath}
\usepackage{changes}
\usepackage{soul, color, xcolor}
\crefname{section}{Sec.}{Secs.}
\Crefname{section}{Section}{Sections}

\definecolor{myblue}{RGB}{200,50,50}

\begin{document}

\title{Quantifying the Upper Limit of Backflash Attack in Quantum Key Distribution}

\author{Jialei Su}
\affiliation{School of Automation, Central South University, Changsha 410083, People's Republic of China}
\affiliation{College of Computer Science and Technology, National University of Defense Technology, Changsha 410073, People's Republic of China}

\author{Junxuan Liu}
\affiliation{College of Computer Science and Technology, National University of Defense Technology, Changsha 410073, People's Republic of China}

\author{Zihao Chen}
\affiliation{College of Computer Science and Technology, National University of Defense Technology, Changsha 410073, People's Republic of China}

\author{Mingyang Zhong}
\affiliation{College of Computer Science and Technology, National University of Defense Technology, Changsha 410073, People's Republic of China}

\author{Qingquan Peng}
\affiliation{College of Computer Science and Technology, National University of Defense Technology, Changsha 410073, People's Republic of China}

\author{Jiangfang Ding}
\affiliation{College of Computer Science and Technology, National University of Defense Technology, Changsha 410073, People's Republic of China}

\author{Yijun Wang}
\email{xxywyj@sina.com}
\affiliation{School of Automation, Central South University, Changsha 410083, People's Republic of China}

\author{Anqi Huang}
\email{angelhuang.hn@gmail.com}
\affiliation{College of Electronic Science and Technology, National University of Defense Technology, Changsha 410073, People's Republic of China}
\affiliation{Center for Cryptologic Research, National University of Defense Technology, Changsha 410073, People's Republic of China}
\affiliation{College of Computer Science and Technology, National University of Defense Technology, Changsha 410073, People's Republic of China}

\author{Ying Guo}
\affiliation{School of Computer, Beijing University of Posts and Telecommunications, Beijing 100876, People's Republic of China}

\begin{abstract}
Quantum key distribution (QKD) provides information-theoretic security grounded in the fundamental laws of physics. Nevertheless, practical imperfections can introduce side channels that expose QKD systems to quantum hacking, especially passive attacks that are inherently difficult to detect. In this study, we experimentally and theoretically investigate the upper limit of the backflash attack—a representative passive side-channel threat. Using a fully equipped fiber-based QKD receiver, we demonstrate the feasibility of the attack and reveal its limited capability in distinguishing quantum states. We further develop a theoretical framework to quantify the maximum distinguishability achievable by an eavesdropper, taking into account the broadband spectral nature of backflash photons. The analysis shows that Eve can extract effective key information from at most 95.7\% of the backflash photons. Based on these findings, we evaluate the secure key rate of a decoy-state BB84 QKD system under backflash attack. Our results provide a quantitative assessment of the vulnerability of QKD systems to backflash emissions and offer a general methodology to evaluate the practical security of QKD systems.
\end{abstract}

\maketitle

\section{Introduction}
\label{sec:intro}

The rapid development of quantum computing has posed a substantial challenge to classical cryptographic algorithms whose security relies on computational complexity~\cite{shor}. Studies have shown that quantum computers have the potential to efficiently break traditional encryption schemes such as RSA~\cite{shao}. Fortunately, quantum key distribution (QKD) offers a fundamentally different approach by leveraging the laws of quantum physics to share symmetric secret keys between two legitimate parties, thereby providing information-theoretic security against eavesdropping~\cite{anquan1,anquan2,anquan3}. However, in practice, realizing an unconditionally secure QKD system remains challenging due to device imperfections, which can be exploited by an eavesdropper (Eve) to gain partial information about the secret key~\cite{loudong,2010hacking,2011hacking,id210,zhongzi,qiang,zhihao2020,2020anastasia,2023anqi,2024qingquan,2024qingquan2}.

Regarding the security of the receiver (Bob) in a prepare-and-measure QKD system, Eve can exploit detector imperfections to extract information about the secret key. Several vulnerabilities in the receiver have been identified, including the blinding attack~\cite{2010hacking,anqi2016,zhihao2020,binwu2022}, the after-gate attack~\cite{houmen}, and the time-shift attack~\cite{shiyi}. 
These attacks on single-photon avalanche detectors (SPADs) are active and require injecting bright light to switch the SPAD's operational state, or using specific methods to intercept and resend quantum states through the quantum channel.
Consequently, Eve’s interventions inevitably cause observable changes in system parameters, such as the quantum bit error rate (QBER), the signal gain, or the detector photocurrent. Changes in system parameters make these types of attacks easy to be detected.

Compared with active attacks, passive attacks are relatively difficult to be detected.
The backflash attack is a typical passive attack targeting SPADs. 
Bob’s SPAD probabilistically emits backflash photons when a detection avalanche occurs~\cite{yingguanganli,yinguangkuanmen,2017,2018}, and these backflash photons transmit backward to the quantum channel and may carry the key information.
Eve can then place a replica of Bob’s measurement devices in the channel and perform the same polarization measurements on the backflash photons to infer the original quantum states.
In a backflash attack, Eve only passively receives the backflash photons without introducing any disturbance to the system parameters.
Therefore, the backflash attack is a security issue that cannot be ignored in practical QKD systems.

The emission of secondary photons in p-n junctions was first reported as early as 1956~\cite{1956}. In recent years, this phenomenon—referred to as backflash—has been revisited in the context of QKD systems. Reference~\cite{2017} analyzed the number of backflash photons emitted by a stand-alone SPAD through optical fiber and proposed a method to measure their spectrum, revealing a broad wavelength distribution. However, this study did not investigate backflash attack within a realistic fiber-based QKD setup. 
For the free-space QKD system, the feasibility of the backflash attack was experimentally demonstrated in~\cite{2018}. It only measured the number of backflash photons in the horizontal ($\ket{H}$) and vertical ($\ket{V}$) polarization states, revealing that Eve's extinction ratio in these two polarization states is limited.
In both free-space and fiber-based QKD systems, Eve’s ability to distinguish quantum states of backflash photons remains uncertain. Specifically, the underlying cause of the limited extinction ratio has not been clarified, leaving the upper bound of Eve’s distinguishing capability unknown. Furthermore, this limitation has not yet been incorporated into security analyses to quantify the corresponding information leakage.

This work investigates the backflash attack in a polarization-encoded, fiber-based QKD system and quantitatively determines its upper limit in leaking secret key information. We experimentally verify Eve’s capability to decode backflash photons and conduct a quantitative analysis of the corresponding photon counts. 
The experimental results reveal that Eve’s decoding ability is intrinsically limited. To further interpret this phenomenon, we develop a general theoretical framework to evaluate the influence of the broadband spectral characteristics of backflash photons on Eve’s decoding capability. The analysis shows that Eve can extract at most 95.7\% of the effective information carried by backflash photons, which is in good agreement with experimental observations. 
Based on both experimental and theoretical findings, we simulated the secure key rate of a decoy-state BB84 QKD system under a backflash attack. The results indicate that the amount of key information accessible to Eve is depends on the setting parameters of SPAD, such as dark count rate and gate width.
By combining experimental verification and theoretical modeling, this study demonstrates comprehensive analysis of the backflash attack on a QKD system, which serves as a general methodology for the security assessment of QKD systems.

The paper is structured as follows. We first introduce the experimental scheme and the testing procedure for the backflash attack in~\cref{sec:setup}. The experimental results are shown in~\cref{sec:results}. \Cref{sec:ER} introduces a theoretical analysis model to verify the limit extinction ratio for Eve decoding the polarization states of backflash photons.
In~\cref{sec:key}, the key rate has been simulated based on the information leakage ratio considering the limited extinction ratio of Eve's decoding.
In~\cref{sec:conclusion}, we conclude our work and present the limitations and potential future work.

\section{Experimental setup and test procedure}
\label{sec:setup}
\subsection{Experimental setup}
\begin{figure}[htbp]
  \includegraphics[width=1\linewidth]{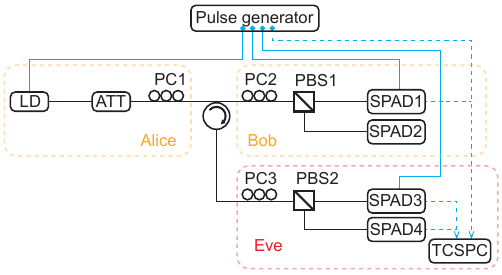}
  \caption{Experimental setup. The black solid lines represent optical fibers, the blue solid lines indicate cables used for triggering devices, and the blue dashed arrows depict the electrical signals transmitted to the time-correlated single photon counting. LD, laser diode; ATT, optical attenuator; PC, polarization controller; CIR, optical circulator; PBS, polarization beam splitter; SPAD, single photon avalanche detector; TCSPC, time-correlated single photon counting.}
  \label{fig:setup}
\end{figure}

This experiment aims to investigate Eve’s capability to acquire secret key information by backflash attack. The eavesdropping scenario is set in a fiber-based QKD system, as shown in~\cref{fig:setup}. On Alice’s side, a laser diode emits optical pulses at a wavelength of $1550~\nano\meter$ with a repetition rate of $100~\kilo\hertz$. The mean photon number is attenuated to $0.5$ photon per pulse, and the pulse width is $1.8~\nano\second$. A polarization controller (PC1) is employed to modulate the polarization states of the emitted photons for encoding.

On Bob’s side, another polarization controller (PC2) compensates for polarization drift in the transmission channel, enabling SPAD1 and SPAD2 to accurately detect the incoming photons. Bob’s detector (IDQ-ID210) operates in external-gating mode, triggered by $100~\kilo\hertz$ electrical pulses. To reduce the dark count rate, the detector dead time is set to $9.8~\micro\second$, resulting in a dark count rate of approximately $50~\hertz$. To enhance the single-photon detection probability, the gate width is set to $15~\nano\second$. It is worth noting that we also characterized other detector parameters to ensure that the tested conditions closely resemble those of practical QKD systems. The corresponding results are presented in Appendix~\ref{sec:appendix1}.

When a detection event occurs in SPAD1 or SPAD2, a backflash photon may be generated with a certain probability. The emitted photon initially exhibits random polarization and carries no decoded information. It is only after passing backward through Bob’s PBS and re-entering the quantum channel that the photon acquires a specific polarization state correlated with the key information. Eve employs an optical circulator to intercept the backflash photons emitted from Bob’s side. The SPADs used by Eve for detection are identical in model to those used by Bob. To maximize the efficiency of decoding the backflash photons, Eve uses another polarization controller (PC3) for polarization calibration. The time-correlated single-photon counting (TCSPC) module is employed to record Eve’s detection events. A pulse generator provides synchronized electrical signals to trigger the laser diode, SPADs, and TCSPC simultaneously.

\subsection{Test procedure}
\label{subsec:test}

To ensure that Eve correctly decodes the quantum states of the backflash photons, Eve’s polarization controller (PC3) must be calibrated before the test. For this purpose, a $1550~\nano\meter$ laser beam is emitted from the position of SPAD1 in the reverse direction of quantum-state transmission, entering the receiver unit through the slow axis of PBS1. PC3 is adjusted to maximize the detection probability of SPAD3 while minimizing that of SPAD4, and the extinction ratio is carefully fine-tuned to approximately $1000$ during the calibration. In this way, the polarization states decoded by Eve based on the backflash photons are closely aligned with those detected by Bob. Specifically, when Bob’s SPAD1 and SPAD2 detect the $\ket{H}$ and $\ket{V}$ states, respectively, Eve’s SPAD3 and SPAD4 correspondingly decode the same polarization states of the backflash photons. The same calibration procedure is applied for the diagonal basis ($\ket{A}$ and $\ket{D}$) as well.

During the experiment, Alice repeatedly emits optical pulses encoded into one of four polarization states ($\ket{H}$, $\ket{V}$, $\ket{A}$, and $\ket{D}$), which are decoded by Bob and detected by his SPADs. Eve’s SPADs operate in two distinct configurations, referred to as Test1 and Test2. In Test1, to observe the temporal distribution of the backflash photons, Eve’s SPADs are set to free-running mode. A $100~\kilo\hertz$ clock signal from the pulse generator is connected to the START channel of the TCSPC, while the detection outputs of Eve’s SPAD3 and SPAD4 (corresponding to $\ket{H}$ and $\ket{V}$, respectively) are connected to the STOP channel.

In Test2, to measure the number of backflash photons, Eve’s SPADs are operated in external-gating mode, where the gate window is configured to fully cover the expected arrival time of the backflash photons. Coincidence counts are recorded within a defined temporal window to exclude backreflection events. In this configuration, the detection signal from Bob’s SPAD1 or SPAD2 is routed to the START channel of the TCSPC, while that from Eve’s SPAD3 or SPAD4 is routed to the STOP channel.

\section{Experimental Results}
\label{sec:results}

\subsection{Verification of attack feasibility}

\begin{figure}[htbp]
	\includegraphics[width=1\linewidth]{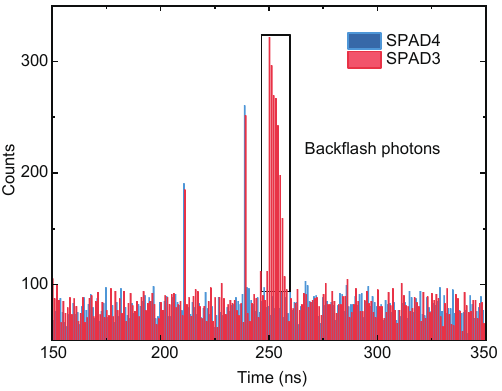}
	\caption{Histogram of the time interval between Alice's laser pulse emission and Eve's detection. When the QKD system transmits the $\ket{H}$ state, the click events of Eve's $\ket{H}$ state detector (SPAD3, red) and $\ket{V}$ state detector (SPAD4, blue) are recorded in the histogram.}
	\label{fig:res}
\end{figure}

In this experiment, Alice repeatedly emits optical pulses with the same polarization state $\ket{H}$, and the click events of SPAD3 (corresponding to $\ket{H}$) and SPAD4 (corresponding to $\ket{V}$) are recorded in a statistical histogram, as shown in Fig.~\ref{fig:res}. The zero point on the horizontal axis corresponds to the moment when Alice emits a pulse. Here, the statistical results are presented over a $200~\nano\second$ time window, where counts outside the identified peaks mainly represent dark counts. These histograms provide a temporal visualization of Eve’s detection events.

The red bars represent the click statistics of SPAD3. Two narrow peaks appearing at earlier times are attributed to backreflections from fiber connectors, while a third peak, observed at approximately 250 ns, corresponds to backflash photons emitted by SPAD1. The temporal width of this peak shows that backflash photons occur probabilistically within about a 10 ns window, which is determined by the gate width of Bob’s detector. The blue bars show the click statistics of SPAD4, which is configured as the orthogonal detector on the same polarization basis as SPAD3. Only the two backreflection peaks are observed for SPAD4, and no statistically significant backflash signal appears near the expected time (around 250 ns). This is because Bob only receives the polarization state $\ket{H}$, so backflash photons carrying the same polarization reach Eve and are detected exclusively by SPAD3, while SPAD4 registers none. The results of Test 1 verify that Eve can infer which of Bob’s detectors clicked by decoding the polarization of the backflash photons, thereby obtaining information correlated with the secret key.

\subsection{Statistical analysis of backflash photons}

To quantify the threat posed by this attack, the key factor is Eve’s ability to distinguish the quantum states carried by the backflash photons. Therefore, it is essential to collect the count of backflash photons detected by each of Eve’s SPADs. In the experiment, Alice is configured to repeatedly prepare the same polarization state. After Bob’s detector registers $5\times10^6$ detection events, the number of corresponding backflash photons detected by Eve is recorded. The results are summarized in Table~\ref{table2}. Here, $C_{E}$ denotes the number of Eve’s detection clicks on the SPAD corresponding to the same quantum state as Bob’s, while $C_{E}^{\perp}$ represents the clicks of the orthogonal detector under the same basis. The extinction ratio on Eve’s side is defined as $ER$.

\begin{table}[h]
	\renewcommand{\arraystretch}{1.5}
	\setlength{\tabcolsep}{10pt}
	\caption{Eve's counts under backflash attack}
	\centering
	\begin{tabular}{lcrrr}
		\hline
		\textbf{Alice's state} & $\ket{H}$ & $\ket{V}$ & $\ket{A}$ & $\ket{D}$  \\
		\hline
		$C_{E}$ & 22526 & 18534 & 18336 & 18854 \\ 
		$C^{\perp}_{E}$ & 796 & 1287 & 741 & 723 \\
		$ER$ & 28.3 & 14.4 & 24.7 & 26.1 \\
		\hline
		\label{table2}
	\end{tabular}
\end{table}

From the test results of the four different polarization states, it can be observed that when Bob emits backflash photons with a specific polarization, the measured $ER$ reaches a maximum of 28.3 across multiple trials, with an average value $\overline{ER}$ of approximately 23.4~\footnote{In practical operation, achieving the ideal $ER$ is challenging due to various factors affecting the measurement. The reduction in $ER$ may result from the non-ideal spectral characteristics of the backflash photons or their broad spectral distribution, which can produce complex responses in wavelength-sensitive components. In addition, imperfect coupling between the single-mode fiber and the polarization-maintaining fiber flange can further contribute to polarization instability. Through continuous optimization and repeated measurements, we acquired a large dataset and selected the optimal $ER$ values for each polarization state, as summarized in Table~I. Notably, the $ER$ obtained for the $\ket{V}$ polarization state is slightly lower than that of the other states. This difference may arise from polarization dispersion and Fresnel effects, which make it more challenging to adjust PC3 to the optimal angle during the decoding of the $\ket{V}$ state.}. However, it is worth emphasizing that during the polarization calibration process, Eve’s extinction ratio reaches approximately 1000 when optical pulses are emitted directly from Bob. The experimental results therefore indicate that the $ER$ values of backflash photons are significantly lower than expected, suggesting that Eve’s ability to distinguish the quantum states of backflash photons is fundamentally limited.

\section{Theoretical Analysis of Eve’s Limited Extinction Ratio}
\label{sec:ER}

The parameter $ER$ reflects Eve’s ability to distinguish quantum states from the backflash photons, where a higher $ER$ indicates that Eve can correctly decode a larger fraction of these photons. In this section, we construct a theoretical simulation to explain the reasons for Eve’s relatively limited $ER$ value in the backflash attack. The detailed modeling process for each optical component is provided in Appendix~\ref{sec:appendix2}.

Previous studies have shown that the spectrum of backflash photons is broadly distributed~\cite{2017}. This characteristic significantly affects Eve’s decoding process, since most optical components used in the QKD system and on Eve’s side exhibit wavelength-dependent responses. When backflash photons with different wavelengths pass through these wavelength-sensitive devices, their polarization states vary accordingly, ultimately limiting the achievable $ER$ for Eve.

\begin{figure}[htbp]
	\includegraphics[width=1\linewidth]{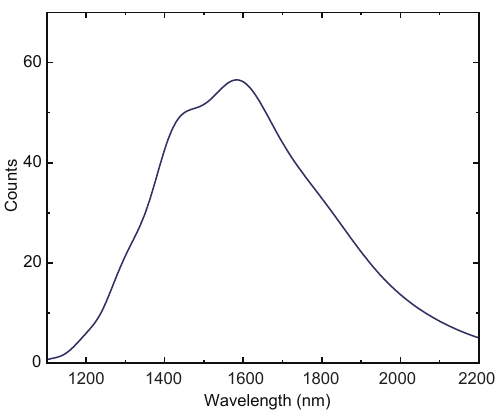}
	\caption {The spectrum of backflash photons by using Ansys Luminescent FDTD simulation platform to simulate the secondary photon emission of avalanche photodiode.}
	\label{fig:spec}
\end{figure}

To quantitatively analyze the influence of the wide spectral distribution of backflash photons on the $ER$ value, it is first necessary to obtain their accurate wavelength distribution. The spectral profile of backflash photons mainly depends on the semiconductor material of the avalanche photodiode. Here, we used the Ansys Luminescent finite-difference time-domain (FDTD) optical simulation platform to model the secondary photon emission of an InGaAs avalanche photodiode. The resulting backflash spectrum, shown in~\cref{fig:spec}, indicates that the photon wavelengths are distributed between $1400~\nano\meter$ and $2000~\nano\meter$, with the highest probability density around $1600~\nano\meter$.

After the backflash photons are emitted from Bob’s SPAD, they pass through PBS1 and PC2, both of which are wavelength-sensitive components. When photons of different wavelengths traverse these devices, their polarization states undergo distinct transformations.
To model the PBS, we plotted the transmission curves of $\ket{H}$-polarized photons as they pass through the PBS’s fast and slow axes, as shown in~\cref{fig:pbs}. The results show that the PBS extinction ratio varies with wavelength mismatch.
For the PC, we computed its Jones matrix based on the actual paddle radius and the number of fiber turns, as given in~\cref{eq.pc}. The simulation results in~\cref{fig:pc} demonstrate that $\ket{H}$-polarized photons at different wavelengths yield distinct output polarization states, indicating that the PC’s effective central wavelength depends on the paddle angles.
Similarly, we modeled the SPAD’s detection efficiency curve as a function of wavelength, shown in~\cref{fig:det}.

We also included the calibration process of PC3, as described in~\cref{subsec:test}, in our simulation. The three paddles of Bob’s PC2 were set to a group of angles $\theta_{B_{1,2,3}}$, corresponding to a configuration where Bob can correctly decode the $\ket{H}$ and $\ket{V}$ states. To simulate Eve’s calibration procedure, each paddle of Eve’s PC3 was scanned in increments of $\pi/50$, resulting in $50^3$ combinations for a full angular sweep. For each of the $50^3$ combinations, the Jones matrices of Bob’s and Eve’s PCs were cascaded. A $\ket{H}$-polarized input was then applied, and projection measurements on the $\ket{H}$/$\ket{V}$ basis yielded probabilities $P_{\ket{H}}$ and $P_{\ket{V}}$. Among all computed configurations, the one giving $P_{\ket{H}} : P_{\ket{V}} \approx 1000 : 1$ (matching the calibration procedure) was selected, and its paddle angles were denoted as $\theta_{E_{1,2,3}}$.

\begin{figure}[htbp]
	\includegraphics[width=1\linewidth]{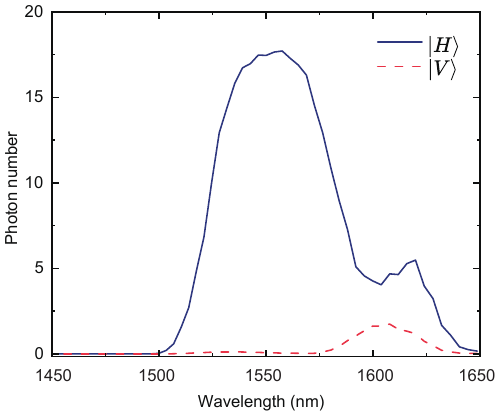}
	\caption {
    The backflash photons with $\ket{H}$ polarization state undergo measurement of their polarization state by Eve after passing through PBS1, PC2, PC3, PBS2 and reaching Eve's SPAD. The blue solid line represents the number of photons decoded as $\ket{H}$ state at different wavelengths, while the red dashed line represents the number of photons decoded as $\ket{V}$ state at different wavelengths.}
	\label{fig:er}
\end{figure}

Subsequently, we simulated the detection of backflash photons. As the photons are emitted from Bob’s SPAD, they pass through PBS1 and PC2 before reaching Eve. After passing through PC3 and PBS3, they are finally detected by Eve’s SPAD. We assume that the backflash photons initially carry $\ket{H}$ polarization before entering Bob’s PC2.
The spectral distribution of the backflash photons was integrated into PC2’s Jones matrix with paddles fixed at $\theta_{B_{1,2,3}}$ for computation. The resulting wavelength-dependent polarization states were then passed through the Jones matrix of PC3 with paddle angles $\theta_{E_{1,2,3}}$. 
The output vectors from PC3 were projected onto the $\ket{H}$/$\ket{V}$ basis, yielding measured probabilities $P'{\ket{H}}$ and $P'{\ket{V}}$ under the backflash attack. Based on the photon number distribution from~\cref{fig:spec} and the PBS/SPAD models described above, the numbers of photons received by Eve for the $\ket{H}$ and $\ket{V}$ states at each wavelength were obtained, as shown in~\cref{fig:er}.
 
Finally, the $ER$ for Eve’s detection of backflash photons was calculated by summing the photon counts for each polarization across all wavelengths. The resulting $ER$ value of 22 closely matches the experimentally measured mean $\overline{ER}$ of 23.4.
When $ER=22$, Eve can correctly decode approximately 95.7\% of the backflash photons. This result defines the upper bound of Eve’s capability to decode backflash photons in the QKD system and confirms the consistency and reliability of our experimental observations.

 \section{Simulation of secure key rate}
 \label{sec:key}

In this section, we estimate the information leakage ratio attributed to the backflash attack based on the experimental data. Using this leakage ratio, we simulate the secure key rate under the impact of the backflash attack. The theoretical simulation is based on the decoy-state BB84 QKD protocol~\cite{wang2005,lo2005,youpian,youpian2}, which is the scheme most commonly implemented in current QKD deployments. In this protocol, in addition to the signal state $\mu$ used for information transmission between Alice and Bob, Alice also emits two decoy states with distinct intensities $\nu_1$ and $\nu_2$ (satisfying $\mu>\nu_1>\nu_2$) to ensure security.

Generally, regarding the weak-coherent pulses emitted by Alice, there is a probability that a single pulse contains multiple photons.
These multi-photon components, denoted as \( \Delta \), are considered to be insecure~\cite{pns}, since Eve is able to learn the information of quantum states from them. 
The fraction of signal states, from which Eve cannot fully obtain information, is represented as \( 1-\Delta \).
The secure key rate in the sifted key is estimated using the inequality as~\cite{2004}
\begin{equation} 
R\geq qQ_\mu[(1-\Delta)-fH_2(\delta)-(1-\Delta)H_2(\frac{\delta}{1-\Delta})].
\label{eq.1}
\end{equation}
Here, $q$ depends on the implementation (1/2 for the BB84 protocol due to Alice and Bob choose different bases in half of the cases, and if one uses the efficient BB84 protocol, $q \approx 1$~\cite{lo2005efficient}).
The $\Delta$ is expressed as \( \Delta = \frac{Q_M}{Q_\mu} \), in which \( Q_M \) denotes the detection probability of Bob's detector responding to multiple-photon pulses, and \( Q_\mu \) represents the overall detection probability of Bob's detector. 
$f$ is the error correction efficiency. 
$\delta$ is the bit error rate.
\( H_2(x) = -x \log_2(x) - (1-x) \log_2(1-x) \) is the binary Shannon entropy function.

If Alice emits multiple photons in one pulse, we pessimistically assume that Eve intercepts at least one photon until the basis is announced and then measures it on the appropriate basis, thereby learning the key bits without introducing any error. 
Within the single-photon component $( 1 - \Delta )$, Bob sacrifices a fraction \( \frac{\delta}{1 - \Delta} \) for privacy amplification. 
We perform a substitution of factors in the equation, let $\frac{\delta}{1-\Delta}=e_1$, $\Delta=\frac{Q_m}{Q_\mu}$, $1-\Delta=\frac{Q_1}{Q_\mu}$, and $\delta=E_\mu$, where $E_\mu$ represents the overall QBER, \(Q_1\) denotes the lower bound for the single-photon gain, and \(e_1\) denotes the upper bound for the single-photon phase error rate. The lower bound for the key rate as
\begin{equation}
R\geq q\{-Q_\mu f(E_\mu)H_2 (E_\mu)+Q_1[1-H_2(e_1)]\}.
\label{equ.2}
\end{equation}
To estimate $ Y_1 $ and $ e_1 $, analytical or numerical tools can be utilized. Here we employ the analytical approach proposed by~Ref.~\cite{youpian} to evaluate the lower bound of \(Y_1\), $Y_0$ and the upper bound of \(e_1\).
\begin{widetext}

\begin{equation}
  	\begin{split}
  Y_1 \geq Y_{1}^{L,v_1,v_2} = \frac{\mu}{\mu v_1 - \mu v_2 - v_1 + v_2}
   \left[ Q_{v_1}e^{ v_1} - Q_{v_2}e^{ v_2} - \frac{v_1^2 - v_2^2}{\mu^2} (Q_{\mu} e^{\mu} - Y_0) \right],
   \end{split}
\end{equation}

\begin{equation}
Y_0 \geq Y_0^L = \max \left\{ \frac{\nu_1 Q_{\nu_2} e^{\nu_2} - \nu_2 Q_{\nu_1} e^{\nu_1}}{\nu_1 - \nu_2}, 0 \right\},
\end{equation}

\begin{equation}
  e_1 \leq e_{1}^{\mu,v_1,v_2} = \frac{E_{v_1} Q_{v_1} e^{v_1} - E_{v_2} Q_{v_2} e^{v_2}}{(\nu_1 - \nu_2) Y_{1}^{L,v_1,v_2}}.
\end{equation}

\end{widetext}

Considering the generation of backflash photons in the SPAD, the probability of backflash emission at Bob’s detector, $P_B$, can be inferred from the number of backflash photons $N_B$ detected by Eve, the detection efficiency of Eve’s detector $\eta_{\text{detE}}$, and the channel transmittance between Bob and Eve $\eta_{\text{chBE}}$, as follows
\begin{equation}
  P_B = \frac{N_B}{N ~\eta_{detE}~ \eta_{\text{chBE}}},
  \label{eq5}
\end{equation}
where $N$ denotes the number of detection events registered by Bob’s SPAD.

Considering the wide-spectrum property of backflash photons, after they pass through the wavelength-sensitive optical device, the polarization drift causes Eve to be unable to fully decode these photons correctly. 
Therefore, we cannot use the total amount of backflash photons to estimate information leakage. 
To reflect Eve's actual decoding capability in a QKD system, we calculate the information leakage rate considering Eve's average extinction ratio $\overline{ER}$ for the backflash photons.
Therefore, the information leakage rate is
\begin{equation}
  P_L =(1-\frac{1}{\overline{ER}+1}) \frac{N_B}{N~ \eta_{detE}~ \eta_{chBE}},
\end{equation}
where \((1-\frac{1}{\overline{ER}+1})\) represents the proportion of backflash photons that Eve can correctly decode. 

Among \( n \) bits of information received by Bob, there is a probability \( P_L \) that a backflash occurs and information leaks. For the backflash attack, we adopt the most pessimistic assumption to estimate the key rate. Specifically, Eve is assumed to be able to capture, store, and measure the quantum states of the backflash photons. Although this is a forward-looking assumption, based on current technology, Eve is also capable of decoding backflash photons. She needs to randomly select bases, just like Bob, and perform projection measurements on the backflash photons. This somewhat limits the amount of information Eve can obtain through backflash attack, but backflash attack currently remains feasible in practical QKD systems. We further consider the most favorable scenario for Eve, in which all backflash photons are successfully detected by her. In this context, the events corresponding to backflash detection are solely attributed to Bob’s single-photon detection events. Therefore, both the components associated with multi-photon emissions and the resulting backflash photons are regarded as potential security vulnerabilities,
\begin{equation}
	\Delta' = \Delta + P_L(1-\Delta).
\end{equation}

By replacing the $\Delta$ in~\cref{eq.1} with $\Delta'$, we can calculate the lower bound of the secure key rate under the backflash attack $R'$.

\begin{table}[h]
	\renewcommand{\arraystretch}{1.5}
	\setlength{\tabcolsep}{12pt}
  	\caption{Experimental parameters used in simulations.}
  	\centering
  	\begin{tabular}{lc}
  		\hline
  		\textbf{Parameter} & \textbf{Value} \\
  		\hline
  		Channel loss coefficient (dB/km) & $\alpha = 0.2$ \\
  		Dark count rate & $P_d =  10^{-4}/gate$ \\
  		Total misalignment error & $e_d = 1\%$ \\
  		Detection efficiency of the SPADs & $\eta_D = 13.9\%$ \\
  		Error correction efficiency & $f = 1.12$ \\
  		\hline
  		\label{table3}
  	\end{tabular}
  	\label{tab:parameters}
\end{table}
  
\begin{figure}[htbp]
  	\includegraphics[width=1\linewidth]{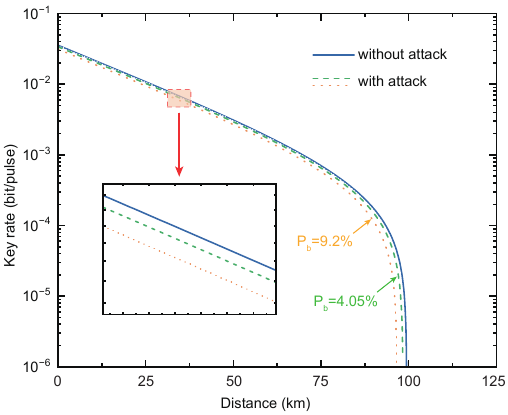}
  	\caption {Simulation results of the decoy-state BB84-QKD protocol. The blue solid line shows the key rate without a backflash attack. The green dashed and yellow dotted lines show the key rates in the presence of a backflash attack, with backflash probabilities of 4.05\% and 9.2\%, respectively.}
  	\label{fig:key}
\end{figure}

Using the experimental parameters listed in~\cref{table3}, the resulting lower bounds of the key rate are depicted in~\cref{fig:key}. 
Here, for each specified distance, we determine the optimal intensities of $\mu_s$, $\nu_1$, and $\nu_2$ to maximize $R$. The intensity of the signal state $\mu_s$ is optimized varying from 0.6 to 0.9 photons per pulse, while the first decoy state $\nu_1$ is optimized within the range of 0.01 to 0.2 photons per pulse. The second decoy state $\nu_2$ is set to the fixed value of $2 \times 10^{-4}$ photons per pulse. According to Eq.~(1), we simulated the secret key rates both with and without the presence of the backflash attack. The solid blue line represents the lower bound of the key rate, $R$, in the absence of the backflash attack. The key rates under attack, denoted as $R'$, were simulated for two scenarios at a repetition frequency of $100~\kilo\hertz$, based on the experimental data presented in Appendix~\ref{sec:appendix1}. When the SPAD gate width was set to $15~\nano\second$, the backflash generation probability was 9.2\%, and the corresponding key rate curve (yellow dotted line) shows that the maximum transmission distance decreases by 2.87\% compared to the no-attack case. When the detector gate width was reduced to $4.99~\nano\second$, the maximum transmission distance decreased by only 1.16\% (green dashed line). These simulation results indicate that the amount of key information accessible to Eve through the backflash attack is extremely limited.

\section{Discussion and Conclusion}
\label{sec:conclusion}
%
In this work, we conducted a comprehensive decoding and quantitative analysis of backflash emissions in a fiber-based QKD system, taking into account the broadband spectral characteristics of spontaneous emission. By experimentally implementing a backflash attack on a polarization-encoded QKD receiver, we found that the number of backflash photons available for Eve’s decoding is limited, primarily due to the wide spectral distribution of the emission. Furthermore, by modeling the physical components within the QKD system, we estimated an upper bound on Eve’s capability to exploit backflash photons, obtaining a theoretical extinction ratio that agrees well with the experimentally measured value. We further evaluated the impact of backflash attack on the secret key generation rate. 
These results demonstrate that the amount of secret key information accessible to Eve through backflash attack is limited, and that the influence of such attacks is dependent on the setting parameters of SPADs used in a QKD system.

Overall, this study quantitatively evaluates the security risk posed by the backflash attack in QKD systems through a combination of experimental verification and theoretical modeling. It not only addresses a gap in existing side-channel security analyses but also establishes an analytical framework for assessing QKD system security against the backflash attack, thereby laying the groundwork for future developments in QKD security evaluation. One of our future objectives is to further reduce the discrepancies between the theoretical analysis of the backflash spectrum and the actual spectral characteristics of backflash photons. Moreover, the current analysis does not distinguish between wavelength components in Eve’s measurement results. If Eve were to spectrally filter the backflash photons and extract key information independently at different wavelengths, she might be able to obtain additional information. Designing and implementing such wavelength-resolved experiments will therefore be an important direction for our future work.

\section*{acknowledgments}
We thank Yaxuan Wang and Tianyi Xing for discussions.
This work was funded by the National Natural Science Foundation of China (No.~62371459) and the Innovation Program for Quantum Science and Technology (2021ZD0300704).

\section*{Data availability}
The data and codes are available from the corresponding author upon request.

\appendix

\section{Factors affecting the generation of backflash photon}
\label{sec:appendix1}
\begin{figure}[htbp]
    \includegraphics[width=1\linewidth]{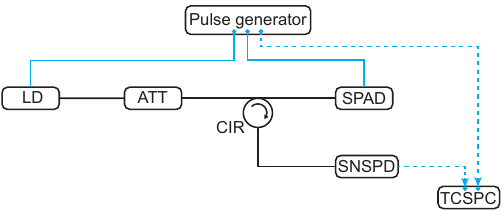}
  	\caption {Test setup for the generation probability of backflash photon. The black solid lines represent optical fibers, the blue solid lines indicate electrical signal for triggering, and the blue dashed arrows depict the electrical signals transmitted to the time-correlated single photon counting (TCSPC). LD, laser diode; ATT, optical attenuator; CIR, optical circulator; SPAD, single photon avalanche detector; SNSPD, superconducting nanowire single-photon detector.}
  	\label{fig:back}
\end{figure}

To explore the factors that may affect the probability of backflash-photon generation, we conduct tests to detect backflash photons under the repetition frequency ranging from $100~\kilo\hertz$ to $100~\mega\hertz$. The testing scheme is shown as~\cref{fig:back}.
A superconducting nanowire single-photon detector (SNSPD) is employed to statistically detect the weak backflash photon generated by a single SPAD. The mean photon number sent by Alice is kept being 0.5 photon per pulse during the testing. The gate width and dead time of the SPAD under test are adjusted accommodating the different repetition frequencies, under the condition that the dark count rate of SPAD is around $10^{-5}/gate$ and the detection efficiency $\eta_{D}$ is around $15\%$. By recording the number of SPAD's click within a fixed time interval of $10~\second$ and counting the number of backflash photons, we calculate the generation probability of the backflash photon $P_b$ as shown in~\cref{table1}.

\begin{table}[h]
	\renewcommand{\arraystretch}{1.5}
	\setlength{\tabcolsep}{4pt}
	\caption{The probability of backflash photon generation under different repetition frequencies and different gate widths. In the table, $f$ denotes the repetition frequency, $GW$ indicates the gate width, and $DT$ signifies the dead time.
}
	\centering
	\begin{tabular}{lccccc}
		\hline
		\textbf {$f$ $($\hertz$)$} & GW $($\nano\second$)$& DT $($\micro\second$)$& DCR $(/gate) $ & $\eta_{D}$ & $P_{b}$ \\
		\hline
            100 k & $15$ & $9.8$ & $5\times 10^{-4}$ & $13.9\%$ & $9.2\%$ \\
		100 k & $10$ & $9.8$ & $9\times 10^{-5}$ & $13.9\%$ & $8.87\%$ \\
            100 k & $4.99$ & $9.8$ & $8\times 10^{-5}$ & $13.4\%$ & $4.05\%$ \\
            500 k & $4.44$ & $3.7$ & $8\times 10^{-5}$ & $14.3\%$ & $3.64\%$ \\
            1 M & $4.3$ & $3.8$ & $7\times 10^{-5}$ & $15.3\%$ & $2.64\%$ \\
            2 M & $3.21$ & $2.7$ & $6\times 10^{-5}$ & $18.3\%$ & $1.65\%$ \\
            5 M & $3.18$ & $2.6$ & $4\times 10^{-5}$ & $15.3\%$ & $1.37\%$ \\
            10 M & $2.97$ & $2.6$ & $4\times 10^{-5}$ & $15.6\%$ & $1.35\%$ \\
            40 M & $2.77$ & $2.6$ & $6\times 10^{-5}$ & $16.9\%$ & $1.32\%$ \\
            80 M & $2.67$ & $2.6$ & $8\times 10^{-5}$ & $15.8\%$ & $1.31\%$ \\
            100 M & $2.55$ & $2.7$ & $8\times 10^{-5}$ & $15.5\%$ & $0.85\%$ \\
            100 M & $1.91$ & $2.1$ & $8\times 10^{-6}$ & $13.1\%$ & $0.17\%$ \\
		\hline
		\label{table1}
	\end{tabular}
\end{table} 

It can be seen from the resting results that, under the similar detection efficiency, the increased repetition frequency results in the decrease of $P_b$, which is mainly due to the change in the gate width. It is shown that the gate width decreasing gradually leads to the generation probability of backflash photon diminishing significantly. When the gate width is reduced to be less than $2~\nano\second$, only few backflash photon emission occurs. This is because a narrower gate width will lead to a shorter avalanche duration, only during which the backflash photons are generated. Then the quenching circuit can release the energy in the APD quickly, thereby reducing the probability of generating backflash photons.

\begin{figure}[htbp]
	\includegraphics[width=1\linewidth]{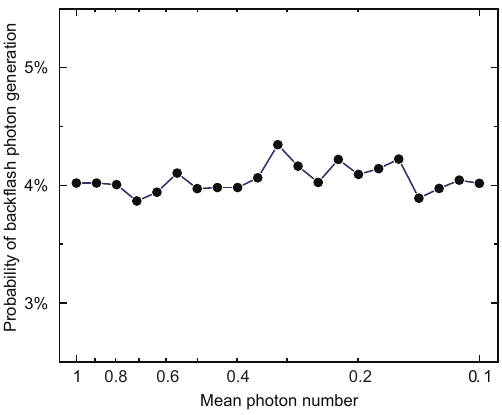}
	\caption {The trend of backflash photon generation probability varying with the mean photon number.}
	\label{fig:u}
\end{figure}

Then, we further investigate the impact of the mean photon number on the generation probability of backflash photon. The parameters of the SPAD are set to be at $100$-$\kilo\hertz$ repetition frequency with gate width of $4.99~\nano\second$ as listed in~\cref{table1}. By varying the mean photon number from 0.1 to 1 photon per pulse, we conducted a statistical analysis of the generation probability of backflash photon. The statistical results are presented in the~\cref{fig:u}, which shows that the probability of backflash photon generation is almost stable under different values of the mean photon number. This is because that, under the condition that the bias voltage and gate width are fixed, the avalanche gain of the detector remains stable, so the backflash detection efficiency at the single-photon level does not significantly vary with the mean photon number.

\bigskip\bigskip

\section{Optical device modeling}
\label{sec:appendix2}

The spectrum of backflash photon is distributed across a certain frequency range. Most of the optical devices through which photons pass are wavelength-sensitive. Simulating by modeling the devices helps to explain the phenomenon of low $ER$ value in Eve's polarization decoding.

We used the Ansys Lumerical FDTD photonic simulation platform for avalanche photodiode simulation. 
The platform provides a secondary photon emission model for InGaAs avalanche photodiodes and allows us to configure relevant parameters, including materials, thickness, and light source.
Then, a monitor is added to detect the secondary photon emission from the avalanche diode and obtain its backflash spectrum.
 The backflash spectrum is shown in~\cref{fig:spec}.

\begin{figure}[h]
	\includegraphics[width=1\linewidth]{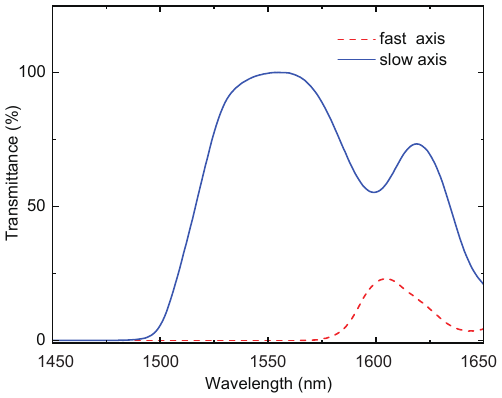}
	\caption {The transmittance of different wavelengths in the two paths of a PBS. The blue solid line represents the transmittance of the $\ket{H}$ state through the slow axis, while the red dashed line represents the transmittance of the $\ket{H}$ state through the fast axis.}
	\label{fig:pbs}
\end{figure}

~\Cref{fig:pbs} shows a PBS with a central wavelength of $1550~\nano\meter$, where the transmittance through the fast axis and the slow axis varies for different wavelengths of photons. Around $1600~\nano\meter$, the transmittance for the fast-axis and slow-axis shows a compensating trend, which, from Eve's perspective, there is a certain probability of decoding into a wrong quantum state, which will increase the bit error for her.

The 3-Paddle Polarization Controller utilizes stress-induced birefringence to create three independent fractional wave plates to alter the polarization in single-mode fiber that is looped around three independent spools(fiber retarders). 
The amount of birefringence induced in the fiber is a function of the fiber cladding diameter, the spool diameter (fixed), the number of fiber loops per spool, and the wavelength of the photons. 
The fast axis of the fiber, which is in the plane of the spool, is adjusted with respect to the transmitted polarization vector by manually rotating the paddles to twist the fiber.
To transform an arbitrary input polarization state into another arbitrary output polarization state, a combination of three paddles (a quarter-wave plate, a half-wave plate, and a quarter-wave plate) can be used. 
The retardance of each paddle may be estimated from the following equation:
\begin{equation}
    \varphi(Radians)=\frac{2\pi^2 a N d^2}{\lambda D},
    \label{eq.paddle}
\end{equation}
\begin{equation}
    \varphi(Waves)=\frac{\pi a N d^2}{\lambda D}.
    \label{eq.paddle1}
\end{equation}
In the formula, $\varphi$ is the phase retardance, $a$ is a constant (for quartz fiber, $a$ equals 0.133), $N$ is the number of fiber loops, $d$ is the diameter of the fiber cladding, $\lambda$ is the wavelength, and $D$ is the diameter of the paddle.
As can be seen from~\cref{eq.paddle} and~\cref{eq.paddle1},  the influence of PC on phase retardance changes depends on the wavelength of the input photons. Assuming that photons with unique initial polarization states but different wavelengths enter the PC, the PC changes the different phase retardances for the various wavelengths of photons. 
List the Jones matrices for the polarization effects of the three paddles:
\begin{widetext}

\begin{equation}
\begin{aligned}
U_{\theta_1}(\varphi_1)=\begin{pmatrix}
	cos\theta_1 & -sin\theta_1 \\
	sin\theta_1 & cos\theta_1 \\
\end{pmatrix}
\begin{pmatrix}
	e^{j\frac{\varphi_1}{2}} & 0 \\
	0 & e^{-j\frac{\varphi_1}{2}} \\
\end{pmatrix}
\begin{pmatrix}
	cos\theta_1 & sin\theta_1 \\
	-sin\theta_1 & cos\theta_1 \\
\end{pmatrix},\\ 
U_{\theta_2}(\varphi_2)=\begin{pmatrix}
	cos\theta_2 & -sin\theta_2 \\
	sin\theta_2 & cos\theta_2 \\
\end{pmatrix}
\begin{pmatrix}
	e^{j\frac{\varphi_2}{2}} & 0 \\
	0 & e^{-j\frac{\varphi_2}{2}} \\
\end{pmatrix}
\begin{pmatrix}
	cos\theta_2 & sin\theta_2 \\
	-sin\theta_2 & cos\theta_2 \\
\end{pmatrix},\\
U_{\theta_3}(\varphi_3)=\begin{pmatrix}
	cos\theta_3 & -sin\theta_3 \\
	sin\theta_3 & cos\theta_3 \\
\end{pmatrix}
\begin{pmatrix}
	e^{j\frac{\varphi_3}{2}} & 0 \\
	0 & e^{-j\frac{\varphi_3}{2}} \\
\end{pmatrix}
\begin{pmatrix}
	cos\theta_3 & sin\theta_3 \\
	-sin\theta_3 & cos\theta_3 \\
\end{pmatrix}.
\end{aligned}
\label{eq.pc}
\end{equation}

\end{widetext}
In each equation, $U$ represents the Jones matrix of the PC's paddle, which is related to the paddle's angle of rotation. Three formulas represent three paddles. By cascading these matrices, the PC's simulation model can be constructed.

\begin{figure}[h]
	\includegraphics[width=1\linewidth]{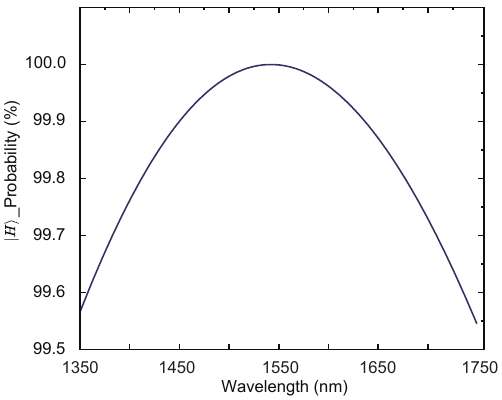}
	\caption {The probability of photons with different wavelengths being measured in the $\ket{H}$ state after passing through the polarization controller with the paddle fixed.}
	\label{fig:pc}
\end{figure}

\begin{figure}[htbp]
	\includegraphics[width=1\linewidth]{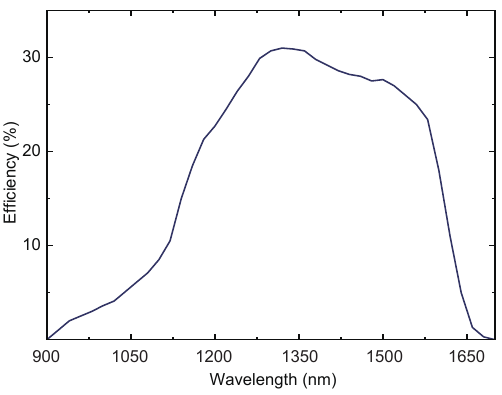}
	\caption {The detection efficiency of a SPAD for photons with different wavelengths.}
	\label{fig:det}
\end{figure}

In the polarization controller model, a $1550~\nano\meter$ photon with arbitrary polarization is used as the input photon, the angles of the paddles are adjusted to ensure that the output polarization state is $\ket{H}$. By fixing the angles of the paddles, we change the wavelength of the input photons and measure the polarization state of the output photons. 
The probability of the measurement results being $\ket{H}$ is calculated as shown in~\cref{fig:pc}.
The horizontal axis represents the wavelength of the input photons, while the vertical axis shows the probability of obtaining the polarization state $\ket{H}$ of the output photons. 
The angle of the PC paddles is set using input photons of $1550~\nano\meter$, so the PC is optimized to adjust the photons in $1550~\nano\meter$ to the state $\ket{H}$. The simulation results confirm the wavelength sensitivity of the PC.

The curve in~\cref{fig:det} represents the detection efficiency of SPAD at various wavelengths. 
For Bob, the efficiency of SPAD corresponds to a specific point, where the detection probability of the $1550~\nano\meter$ photons is 25\%. For Eve, the SPAD efficiency corresponds to a segment of the curve.
As backflash photons of different wavelengths reach Eve's detector, the efficiency of detecting varies.


\begin{thebibliography}{33}%
\makeatletter
\providecommand \@ifxundefined [1]{%
 \@ifx{#1\undefined}
}%
\providecommand \@ifnum [1]{%
 \ifnum #1\expandafter \@firstoftwo
 \else \expandafter \@secondoftwo
 \fi
}%
\providecommand \@ifx [1]{%
 \ifx #1\expandafter \@firstoftwo
 \else \expandafter \@secondoftwo
 \fi
}%
\providecommand \natexlab [1]{#1}%
\providecommand \enquote  [1]{``#1''}%
\providecommand \bibnamefont  [1]{#1}%
\providecommand \bibfnamefont [1]{#1}%
\providecommand \citenamefont [1]{#1}%
\providecommand \href@noop [0]{\@secondoftwo}%
\providecommand \href [0]{\begingroup \@sanitize@url \@href}%
\providecommand \@href[1]{\@@startlink{#1}\@@href}%
\providecommand \@@href[1]{\endgroup#1\@@endlink}%
\providecommand \@sanitize@url [0]{\catcode `\\12\catcode `\$12\catcode
  `\&12\catcode `\#12\catcode `\^12\catcode `\_12\catcode `\%12\relax}%
\providecommand \@@startlink[1]{}%
\providecommand \@@endlink[0]{}%
\providecommand \url  [0]{\begingroup\@sanitize@url \@url }%
\providecommand \@url [1]{\endgroup\@href {#1}{\urlprefix }}%
\providecommand \urlprefix  [0]{URL }%
\providecommand \Eprint [0]{\href }%
\providecommand \doibase [0]{https://doi.org/}%
\providecommand \selectlanguage [0]{\@gobble}%
\providecommand \bibinfo  [0]{\@secondoftwo}%
\providecommand \bibfield  [0]{\@secondoftwo}%
\providecommand \translation [1]{[#1]}%
\providecommand \BibitemOpen [0]{}%
\providecommand \bibitemStop [0]{}%
\providecommand \bibitemNoStop [0]{.\EOS\space}%
\providecommand \EOS [0]{\spacefactor3000\relax}%
\providecommand \BibitemShut  [1]{\csname bibitem#1\endcsname}%
\let\auto@bib@innerbib\@empty
\bibitem [{\citenamefont {Shor}(1999)}]{shor}%
  \BibitemOpen
  \bibfield  {author} {\bibinfo {author} {\bibfnamefont {P.~W.}\ \bibnamefont
  {Shor}},\ }\bibfield  {title} {\bibinfo {title} {Polynomial-time algorithms
  for prime factorization and discrete logarithms on a quantum computer},\
  }\href {https://doi.org/10.1137/S0036144598347011} {\bibfield  {journal}
  {\bibinfo  {journal} {SIAM Review}\ }\textbf {\bibinfo {volume} {41}},\
  \bibinfo {pages} {303} (\bibinfo {year} {1999})}\BibitemShut {NoStop}%
\bibitem [{\citenamefont {Shor}(1994)}]{shao}%
  \BibitemOpen
  \bibfield  {author} {\bibinfo {author} {\bibfnamefont {P.}~\bibnamefont
  {Shor}},\ }\bibfield  {title} {\bibinfo {title} {Algorithms for quantum
  computation: discrete logarithms and factoring},\ }in\ \href
  {https://doi.org/10.1109/SFCS.1994.365700} {\emph {\bibinfo {booktitle}
  {Proceedings 35th Annual Symposium on Foundations of Computer Science}}}\
  (\bibinfo {year} {1994})\ pp.\ \bibinfo {pages} {124--134}\BibitemShut
  {NoStop}%
\bibitem [{\citenamefont {H.~Zbinden}\ and\ \citenamefont
  {Tittel}(2002)}]{anquan1}%
  \BibitemOpen
  \bibfield  {author} {\bibinfo {author} {\bibfnamefont {G.~R. D.~S.}\
  \bibnamefont {H.~Zbinden}, \bibfnamefont {N.~Gisin}}\ and\ \bibinfo {author}
  {\bibfnamefont {W.}~\bibnamefont {Tittel}},\ }\bibfield  {title} {\bibinfo
  {title} {Experimental quantum communication},\ }in\ \href
  {https://doi.org/10.3254/978-1-61499-004-8-217} {\emph {\bibinfo {booktitle}
  {Experimental Quantum Computation and Information}}}\ (\bibinfo  {publisher}
  {IOS Press},\ \bibinfo {year} {2002})\ pp.\ \bibinfo {pages}
  {217--232}\BibitemShut {NoStop}%
\bibitem [{\citenamefont {Scarani}\ \emph {et~al.}(2009)\citenamefont
  {Scarani}, \citenamefont {Bechmann-Pasquinucci}, \citenamefont {Cerf},
  \citenamefont {Du\ifmmode~\check{s}\else \v{s}\fi{}ek}, \citenamefont
  {L\"utkenhaus},\ and\ \citenamefont {Peev}}]{anquan2}%
  \BibitemOpen
  \bibfield  {author} {\bibinfo {author} {\bibfnamefont {V.}~\bibnamefont
  {Scarani}}, \bibinfo {author} {\bibfnamefont {H.}~\bibnamefont
  {Bechmann-Pasquinucci}}, \bibinfo {author} {\bibfnamefont {N.~J.}\
  \bibnamefont {Cerf}}, \bibinfo {author} {\bibfnamefont {M.}~\bibnamefont
  {Du\ifmmode~\check{s}\else \v{s}\fi{}ek}}, \bibinfo {author} {\bibfnamefont
  {N.}~\bibnamefont {L\"utkenhaus}},\ and\ \bibinfo {author} {\bibfnamefont
  {M.}~\bibnamefont {Peev}},\ }\bibfield  {title} {\bibinfo {title} {The
  security of practical quantum key distribution},\ }\href
  {https://doi.org/10.1103/RevModPhys.81.1301} {\bibfield  {journal} {\bibinfo
  {journal} {Rev. Mod. Phys.}\ }\textbf {\bibinfo {volume} {81}},\ \bibinfo
  {pages} {1301} (\bibinfo {year} {2009})}\BibitemShut {NoStop}%
\bibitem [{\citenamefont {Lo}\ \emph {et~al.}(2014)\citenamefont {Lo},
  \citenamefont {Curty},\ and\ \citenamefont {Tamaki}}]{anquan3}%
  \BibitemOpen
  \bibfield  {author} {\bibinfo {author} {\bibfnamefont {H.-K.}\ \bibnamefont
  {Lo}}, \bibinfo {author} {\bibfnamefont {M.}~\bibnamefont {Curty}},\ and\
  \bibinfo {author} {\bibfnamefont {K.}~\bibnamefont {Tamaki}},\ }\bibfield
  {title} {\bibinfo {title} {Secure quantum key distribution},\ }\href
  {https://doi.org/10.1038/nphoton.2014.149} {\bibfield  {journal} {\bibinfo
  {journal} {Nat. Photonics}\ }\textbf {\bibinfo {volume} {8}},\ \bibinfo
  {pages} {595} (\bibinfo {year} {2014})}\BibitemShut {NoStop}%
\bibitem [{\citenamefont {Makarov}\ \emph {et~al.}(2006)\citenamefont
  {Makarov}, \citenamefont {Anisimov},\ and\ \citenamefont {Skaar}}]{loudong}%
  \BibitemOpen
  \bibfield  {author} {\bibinfo {author} {\bibfnamefont {V.}~\bibnamefont
  {Makarov}}, \bibinfo {author} {\bibfnamefont {A.}~\bibnamefont {Anisimov}},\
  and\ \bibinfo {author} {\bibfnamefont {J.}~\bibnamefont {Skaar}},\ }\bibfield
   {title} {\bibinfo {title} {Effects of detector efficiency mismatch on
  security of quantum cryptosystems},\ }\href
  {https://doi.org/10.1103/PhysRevA.74.022313} {\bibfield  {journal} {\bibinfo
  {journal} {Phys. Rev. A}\ }\textbf {\bibinfo {volume} {74}},\ \bibinfo
  {pages} {022313} (\bibinfo {year} {2006})}\BibitemShut {NoStop}%
\bibitem [{\citenamefont {Lydersen}\ \emph {et~al.}(2010)\citenamefont
  {Lydersen}, \citenamefont {Wiechers}, \citenamefont {Wittmann}, \citenamefont
  {Elser}, \citenamefont {Skaar},\ and\ \citenamefont {Makarov}}]{2010hacking}%
  \BibitemOpen
  \bibfield  {author} {\bibinfo {author} {\bibfnamefont {L.}~\bibnamefont
  {Lydersen}}, \bibinfo {author} {\bibfnamefont {C.}~\bibnamefont {Wiechers}},
  \bibinfo {author} {\bibfnamefont {C.}~\bibnamefont {Wittmann}}, \bibinfo
  {author} {\bibfnamefont {D.}~\bibnamefont {Elser}}, \bibinfo {author}
  {\bibfnamefont {J.}~\bibnamefont {Skaar}},\ and\ \bibinfo {author}
  {\bibfnamefont {V.}~\bibnamefont {Makarov}},\ }\bibfield  {title} {\bibinfo
  {title} {Hacking commercial quantum cryptography systems by tailored bright
  illumination},\ }\href {https://doi.org/10.1038/nphoton.2010.214} {\bibfield
  {journal} {\bibinfo  {journal} {Nat. Photonics}\ }\textbf {\bibinfo {volume}
  {4}},\ \bibinfo {pages} {686} (\bibinfo {year} {2010})}\BibitemShut {NoStop}%
\bibitem [{\citenamefont {Gerhardt}\ \emph {et~al.}(2011)\citenamefont
  {Gerhardt}, \citenamefont {Liu}, \citenamefont {Lamas-Linares}, \citenamefont
  {Skaar}, \citenamefont {Kurtsiefer},\ and\ \citenamefont
  {Makarov}}]{2011hacking}%
  \BibitemOpen
  \bibfield  {author} {\bibinfo {author} {\bibfnamefont {I.}~\bibnamefont
  {Gerhardt}}, \bibinfo {author} {\bibfnamefont {Q.}~\bibnamefont {Liu}},
  \bibinfo {author} {\bibfnamefont {A.}~\bibnamefont {Lamas-Linares}}, \bibinfo
  {author} {\bibfnamefont {J.}~\bibnamefont {Skaar}}, \bibinfo {author}
  {\bibfnamefont {C.}~\bibnamefont {Kurtsiefer}},\ and\ \bibinfo {author}
  {\bibfnamefont {V.}~\bibnamefont {Makarov}},\ }\bibfield  {title} {\bibinfo
  {title} {Full-field implementation of a perfect eavesdropper on a quantum
  cryptography system},\ }\href {https://doi.org/10.1038/ncomms1348} {\bibfield
   {journal} {\bibinfo  {journal} {Nat. Commun}\ }\textbf {\bibinfo {volume}
  {2}},\ \bibinfo {pages} {349} (\bibinfo {year} {2011})}\BibitemShut {NoStop}%
\bibitem [{\citenamefont {Chistiakov}\ \emph {et~al.}(2019)\citenamefont
  {Chistiakov}, \citenamefont {Huang}, \citenamefont {Egorov},\ and\
  \citenamefont {Makarov}}]{id210}%
  \BibitemOpen
  \bibfield  {author} {\bibinfo {author} {\bibfnamefont {V.}~\bibnamefont
  {Chistiakov}}, \bibinfo {author} {\bibfnamefont {A.}~\bibnamefont {Huang}},
  \bibinfo {author} {\bibfnamefont {V.}~\bibnamefont {Egorov}},\ and\ \bibinfo
  {author} {\bibfnamefont {V.}~\bibnamefont {Makarov}},\ }\bibfield  {title}
  {\bibinfo {title} {Controlling single-photon detector id210 with bright
  light},\ }\href {https://doi.org/10.1364/OE.27.032253} {\bibfield  {journal}
  {\bibinfo  {journal} {Opt. Express}\ }\textbf {\bibinfo {volume} {27}},\
  \bibinfo {pages} {32253} (\bibinfo {year} {2019})}\BibitemShut {NoStop}%
\bibitem [{\citenamefont {Huang}\ \emph {et~al.}(2019)\citenamefont {Huang},
  \citenamefont {Navarrete}, \citenamefont {Sun}, \citenamefont {Chaiwongkhot},
  \citenamefont {Curty},\ and\ \citenamefont {Makarov}}]{zhongzi}%
  \BibitemOpen
  \bibfield  {author} {\bibinfo {author} {\bibfnamefont {A.}~\bibnamefont
  {Huang}}, \bibinfo {author} {\bibfnamefont {A.}~\bibnamefont {Navarrete}},
  \bibinfo {author} {\bibfnamefont {S.-H.}\ \bibnamefont {Sun}}, \bibinfo
  {author} {\bibfnamefont {P.}~\bibnamefont {Chaiwongkhot}}, \bibinfo {author}
  {\bibfnamefont {M.}~\bibnamefont {Curty}},\ and\ \bibinfo {author}
  {\bibfnamefont {V.}~\bibnamefont {Makarov}},\ }\bibfield  {title} {\bibinfo
  {title} {Laser-seeding attack in quantum key distribution},\ }\href
  {https://doi.org/10.1103/PhysRevApplied.12.064043} {\bibfield  {journal}
  {\bibinfo  {journal} {Phys. Rev. Appl.}\ }\textbf {\bibinfo {volume} {12}},\
  \bibinfo {pages} {064043} (\bibinfo {year} {2019})}\BibitemShut {NoStop}%
\bibitem [{\citenamefont {Huang}\ \emph {et~al.}(2020)\citenamefont {Huang},
  \citenamefont {Li}, \citenamefont {Egorov}, \citenamefont {Tchouragoulov},
  \citenamefont {Kumar},\ and\ \citenamefont {Makarov}}]{qiang}%
  \BibitemOpen
  \bibfield  {author} {\bibinfo {author} {\bibfnamefont {A.}~\bibnamefont
  {Huang}}, \bibinfo {author} {\bibfnamefont {R.}~\bibnamefont {Li}}, \bibinfo
  {author} {\bibfnamefont {V.}~\bibnamefont {Egorov}}, \bibinfo {author}
  {\bibfnamefont {S.}~\bibnamefont {Tchouragoulov}}, \bibinfo {author}
  {\bibfnamefont {K.}~\bibnamefont {Kumar}},\ and\ \bibinfo {author}
  {\bibfnamefont {V.}~\bibnamefont {Makarov}},\ }\bibfield  {title} {\bibinfo
  {title} {Laser-damage attack against optical attenuators in quantum key
  distribution},\ }\href {https://doi.org/10.1103/PhysRevApplied.13.034017}
  {\bibfield  {journal} {\bibinfo  {journal} {Phys. Rev. Appl.}\ }\textbf
  {\bibinfo {volume} {13}},\ \bibinfo {pages} {034017} (\bibinfo {year}
  {2020})}\BibitemShut {NoStop}%
\bibitem [{\citenamefont {Wu}\ \emph {et~al.}(2020)\citenamefont {Wu},
  \citenamefont {Huang}, \citenamefont {Chen}, \citenamefont {Sun},
  \citenamefont {Ding}, \citenamefont {Qiang}, \citenamefont {Fu},
  \citenamefont {Xu},\ and\ \citenamefont {Wu}}]{zhihao2020}%
  \BibitemOpen
  \bibfield  {author} {\bibinfo {author} {\bibfnamefont {Z.}~\bibnamefont
  {Wu}}, \bibinfo {author} {\bibfnamefont {A.}~\bibnamefont {Huang}}, \bibinfo
  {author} {\bibfnamefont {H.}~\bibnamefont {Chen}}, \bibinfo {author}
  {\bibfnamefont {S.-H.}\ \bibnamefont {Sun}}, \bibinfo {author} {\bibfnamefont
  {J.}~\bibnamefont {Ding}}, \bibinfo {author} {\bibfnamefont {X.}~\bibnamefont
  {Qiang}}, \bibinfo {author} {\bibfnamefont {X.}~\bibnamefont {Fu}}, \bibinfo
  {author} {\bibfnamefont {P.}~\bibnamefont {Xu}},\ and\ \bibinfo {author}
  {\bibfnamefont {J.}~\bibnamefont {Wu}},\ }\bibfield  {title} {\bibinfo
  {title} {Hacking single-photon avalanche detectors in quantum key
  distribution via pulse illumination},\ }\href
  {https://doi.org/10.1364/OE.397962} {\bibfield  {journal} {\bibinfo
  {journal} {Opt. Express}\ }\textbf {\bibinfo {volume} {28}},\ \bibinfo
  {pages} {25574} (\bibinfo {year} {2020})}\BibitemShut {NoStop}%
\bibitem [{\citenamefont {Ponosova}\ \emph {et~al.}(2022)\citenamefont
  {Ponosova}, \citenamefont {Ruzhitskaya}, \citenamefont {Chaiwongkhot},
  \citenamefont {Egorov}, \citenamefont {Makarov},\ and\ \citenamefont
  {Huang}}]{2020anastasia}%
  \BibitemOpen
  \bibfield  {author} {\bibinfo {author} {\bibfnamefont {A.}~\bibnamefont
  {Ponosova}}, \bibinfo {author} {\bibfnamefont {D.}~\bibnamefont
  {Ruzhitskaya}}, \bibinfo {author} {\bibfnamefont {P.}~\bibnamefont
  {Chaiwongkhot}}, \bibinfo {author} {\bibfnamefont {V.}~\bibnamefont
  {Egorov}}, \bibinfo {author} {\bibfnamefont {V.}~\bibnamefont {Makarov}},\
  and\ \bibinfo {author} {\bibfnamefont {A.}~\bibnamefont {Huang}},\ }\bibfield
   {title} {\bibinfo {title} {Protecting fiber-optic quantum key distribution
  sources against light-injection attacks},\ }\href
  {https://doi.org/10.1103/PRXQuantum.3.040307} {\bibfield  {journal} {\bibinfo
   {journal} {PRX Quantum}\ }\textbf {\bibinfo {volume} {3}},\ \bibinfo {pages}
  {040307} (\bibinfo {year} {2022})}\BibitemShut {NoStop}%
\bibitem [{\citenamefont {Huang}\ \emph {et~al.}(2023)\citenamefont {Huang},
  \citenamefont {Mizutani}, \citenamefont {Lo}, \citenamefont {Makarov},\ and\
  \citenamefont {Tamaki}}]{2023anqi}%
  \BibitemOpen
  \bibfield  {author} {\bibinfo {author} {\bibfnamefont {A.}~\bibnamefont
  {Huang}}, \bibinfo {author} {\bibfnamefont {A.}~\bibnamefont {Mizutani}},
  \bibinfo {author} {\bibfnamefont {H.-K.}\ \bibnamefont {Lo}}, \bibinfo
  {author} {\bibfnamefont {V.}~\bibnamefont {Makarov}},\ and\ \bibinfo {author}
  {\bibfnamefont {K.}~\bibnamefont {Tamaki}},\ }\bibfield  {title} {\bibinfo
  {title} {Characterization of state-preparation uncertainty in quantum key
  distribution},\ }\href {https://doi.org/10.1103/PhysRevApplied.19.014048}
  {\bibfield  {journal} {\bibinfo  {journal} {Phys. Rev. Appl.}\ }\textbf
  {\bibinfo {volume} {19}},\ \bibinfo {pages} {014048} (\bibinfo {year}
  {2023})}\BibitemShut {NoStop}%
\bibitem [{\citenamefont {Peng}\ \emph {et~al.}(2024)\citenamefont {Peng},
  \citenamefont {Gao}, \citenamefont {Zaitsev}, \citenamefont {Wang},
  \citenamefont {Ding}, \citenamefont {Liu}, \citenamefont {Liao},
  \citenamefont {Guo}, \citenamefont {Huang},\ and\ \citenamefont
  {Wu}}]{2024qingquan}%
  \BibitemOpen
  \bibfield  {author} {\bibinfo {author} {\bibfnamefont {Q.}~\bibnamefont
  {Peng}}, \bibinfo {author} {\bibfnamefont {B.}~\bibnamefont {Gao}}, \bibinfo
  {author} {\bibfnamefont {K.}~\bibnamefont {Zaitsev}}, \bibinfo {author}
  {\bibfnamefont {D.}~\bibnamefont {Wang}}, \bibinfo {author} {\bibfnamefont
  {J.}~\bibnamefont {Ding}}, \bibinfo {author} {\bibfnamefont {Y.}~\bibnamefont
  {Liu}}, \bibinfo {author} {\bibfnamefont {Q.}~\bibnamefont {Liao}}, \bibinfo
  {author} {\bibfnamefont {Y.}~\bibnamefont {Guo}}, \bibinfo {author}
  {\bibfnamefont {A.}~\bibnamefont {Huang}},\ and\ \bibinfo {author}
  {\bibfnamefont {J.}~\bibnamefont {Wu}},\ }\bibfield  {title} {\bibinfo
  {title} {Security boundaries of an optical-power limiter for protecting
  quantum-key-distribution systems},\ }\href
  {https://doi.org/10.1103/PhysRevApplied.21.014026} {\bibfield  {journal}
  {\bibinfo  {journal} {Phys. Rev. Appl.}\ }\textbf {\bibinfo {volume} {21}},\
  \bibinfo {pages} {014026} (\bibinfo {year} {2024})}\BibitemShut {NoStop}%
\bibitem [{\citenamefont {{Peng}}\ \emph {et~al.}()\citenamefont {{Peng}},
  \citenamefont {{Chen}}, \citenamefont {{Xing}}, \citenamefont {{Wang}},
  \citenamefont {{Wang}}, \citenamefont {{Liu}},\ and\ \citenamefont
  {{Huang}}}]{2024qingquan2}%
  \BibitemOpen
  \bibfield  {author} {\bibinfo {author} {\bibfnamefont {Q.}~\bibnamefont
  {{Peng}}}, \bibinfo {author} {\bibfnamefont {J.-P.}\ \bibnamefont {{Chen}}},
  \bibinfo {author} {\bibfnamefont {T.}~\bibnamefont {{Xing}}}, \bibinfo
  {author} {\bibfnamefont {D.}~\bibnamefont {{Wang}}}, \bibinfo {author}
  {\bibfnamefont {Y.}~\bibnamefont {{Wang}}}, \bibinfo {author} {\bibfnamefont
  {Y.}~\bibnamefont {{Liu}}},\ and\ \bibinfo {author} {\bibfnamefont
  {A.}~\bibnamefont {{Huang}}},\ }\href@noop {} {\bibinfo {title} {{Practical
  security of twin-field quantum key distribution under wavelength-switching
  attack}}},\ \Eprint {https://arxiv.org/abs/2408.09318} {arXiv:2408.09318
  [quant-ph]} \BibitemShut {NoStop}%
\bibitem [{\citenamefont {Huang}\ \emph {et~al.}(2016)\citenamefont {Huang},
  \citenamefont {Sajeed}, \citenamefont {Chaiwongkhot}, \citenamefont
  {Soucarros}, \citenamefont {Legré},\ and\ \citenamefont
  {Makarov}}]{anqi2016}%
  \BibitemOpen
  \bibfield  {author} {\bibinfo {author} {\bibfnamefont {A.}~\bibnamefont
  {Huang}}, \bibinfo {author} {\bibfnamefont {S.}~\bibnamefont {Sajeed}},
  \bibinfo {author} {\bibfnamefont {P.}~\bibnamefont {Chaiwongkhot}}, \bibinfo
  {author} {\bibfnamefont {M.}~\bibnamefont {Soucarros}}, \bibinfo {author}
  {\bibfnamefont {M.}~\bibnamefont {Legré}},\ and\ \bibinfo {author}
  {\bibfnamefont {V.}~\bibnamefont {Makarov}},\ }\bibfield  {title} {\bibinfo
  {title} {Testing random-detector-efficiency countermeasure in a commercial
  system reveals a breakable unrealistic assumption},\ }\href
  {https://doi.org/10.1109/JQE.2016.2611443} {\bibfield  {journal} {\bibinfo
  {journal} {IEEE J. of Quantum Electronics}\ }\textbf {\bibinfo {volume}
  {52}},\ \bibinfo {pages} {1} (\bibinfo {year} {2016})}\BibitemShut {NoStop}%
\bibitem [{\citenamefont {Gao}\ \emph {et~al.}(2022)\citenamefont {Gao},
  \citenamefont {Wu}, \citenamefont {Shi}, \citenamefont {Liu}, \citenamefont
  {Wang}, \citenamefont {Yu}, \citenamefont {Huang},\ and\ \citenamefont
  {Wu}}]{binwu2022}%
  \BibitemOpen
  \bibfield  {author} {\bibinfo {author} {\bibfnamefont {B.}~\bibnamefont
  {Gao}}, \bibinfo {author} {\bibfnamefont {Z.}~\bibnamefont {Wu}}, \bibinfo
  {author} {\bibfnamefont {W.}~\bibnamefont {Shi}}, \bibinfo {author}
  {\bibfnamefont {Y.}~\bibnamefont {Liu}}, \bibinfo {author} {\bibfnamefont
  {D.}~\bibnamefont {Wang}}, \bibinfo {author} {\bibfnamefont {C.}~\bibnamefont
  {Yu}}, \bibinfo {author} {\bibfnamefont {A.}~\bibnamefont {Huang}},\ and\
  \bibinfo {author} {\bibfnamefont {J.}~\bibnamefont {Wu}},\ }\bibfield
  {title} {\bibinfo {title} {Ability of strong-pulse illumination to hack
  self-differencing avalanche photodiode detectors in a high-speed
  quantum-key-distribution system},\ }\href
  {https://doi.org/10.1103/PhysRevA.106.033713} {\bibfield  {journal} {\bibinfo
   {journal} {Phys. Rev. A}\ }\textbf {\bibinfo {volume} {106}},\ \bibinfo
  {pages} {033713} (\bibinfo {year} {2022})}\BibitemShut {NoStop}%
\bibitem [{\citenamefont {Wiechers}\ \emph {et~al.}(2011)\citenamefont
  {Wiechers}, \citenamefont {Lydersen}, \citenamefont {Wittmann}, \citenamefont
  {Elser}, \citenamefont {Skaar}, \citenamefont {Marquardt}, \citenamefont
  {Makarov},\ and\ \citenamefont {Leuchs}}]{houmen}%
  \BibitemOpen
  \bibfield  {author} {\bibinfo {author} {\bibfnamefont {C.}~\bibnamefont
  {Wiechers}}, \bibinfo {author} {\bibfnamefont {L.}~\bibnamefont {Lydersen}},
  \bibinfo {author} {\bibfnamefont {C.}~\bibnamefont {Wittmann}}, \bibinfo
  {author} {\bibfnamefont {D.}~\bibnamefont {Elser}}, \bibinfo {author}
  {\bibfnamefont {J.}~\bibnamefont {Skaar}}, \bibinfo {author} {\bibfnamefont
  {C.}~\bibnamefont {Marquardt}}, \bibinfo {author} {\bibfnamefont
  {V.}~\bibnamefont {Makarov}},\ and\ \bibinfo {author} {\bibfnamefont
  {G.}~\bibnamefont {Leuchs}},\ }\bibfield  {title} {\bibinfo {title}
  {After-gate attack on a quantum cryptosystem},\ }\href
  {https://doi.org/10.1088/1367-2630/13/1/013043} {\bibfield  {journal}
  {\bibinfo  {journal} {New J. Phys}\ }\textbf {\bibinfo {volume} {13}},\
  \bibinfo {pages} {013043} (\bibinfo {year} {2011})}\BibitemShut {NoStop}%
\bibitem [{\citenamefont {Zhao}\ \emph {et~al.}(2008)\citenamefont {Zhao},
  \citenamefont {Fung}, \citenamefont {Qi}, \citenamefont {Chen},\ and\
  \citenamefont {Lo}}]{shiyi}%
  \BibitemOpen
  \bibfield  {author} {\bibinfo {author} {\bibfnamefont {Y.}~\bibnamefont
  {Zhao}}, \bibinfo {author} {\bibfnamefont {C.-H.~F.}\ \bibnamefont {Fung}},
  \bibinfo {author} {\bibfnamefont {B.}~\bibnamefont {Qi}}, \bibinfo {author}
  {\bibfnamefont {C.}~\bibnamefont {Chen}},\ and\ \bibinfo {author}
  {\bibfnamefont {H.-K.}\ \bibnamefont {Lo}},\ }\bibfield  {title} {\bibinfo
  {title} {Quantum hacking: Experimental demonstration of time-shift attack
  against practical quantum-key-distribution systems},\ }\href
  {https://doi.org/10.1103/PhysRevA.78.042333} {\bibfield  {journal} {\bibinfo
  {journal} {Phys. Rev. A}\ }\textbf {\bibinfo {volume} {78}},\ \bibinfo
  {pages} {042333} (\bibinfo {year} {2008})}\BibitemShut {NoStop}%
\bibitem [{\citenamefont {Meda}\ \emph {et~al.}(2018)\citenamefont {Meda},
  \citenamefont {Degiovanni}, \citenamefont {Tosi}, \citenamefont {Yuan},
  \citenamefont {Brida},\ and\ \citenamefont {Genovese}}]{yingguanganli}%
  \BibitemOpen
  \bibfield  {author} {\bibinfo {author} {\bibfnamefont {A.}~\bibnamefont
  {Meda}}, \bibinfo {author} {\bibfnamefont {I.}~\bibnamefont {Degiovanni}},
  \bibinfo {author} {\bibfnamefont {A.}~\bibnamefont {Tosi}}, \bibinfo {author}
  {\bibfnamefont {Z.}~\bibnamefont {Yuan}}, \bibinfo {author} {\bibfnamefont
  {G.}~\bibnamefont {Brida}},\ and\ \bibinfo {author} {\bibfnamefont
  {M.}~\bibnamefont {Genovese}},\ }\bibfield  {title} {\bibinfo {title}
  {Quantum key distribution security threat: the backflash light case},\ }in\
  \href {https://doi.org/10.1117/12.2307704} {\emph {\bibinfo {booktitle}
  {Quantum Technologies 2018}}},\ Vol.\ \bibinfo {volume} {10674}\ (\bibinfo
  {organization} {SPIE},\ \bibinfo {year} {2018})\ pp.\ \bibinfo {pages}
  {138--144}\BibitemShut {NoStop}%
\bibitem [{\citenamefont {Koehler-Sidki}\ \emph {et~al.}(2020)\citenamefont
  {Koehler-Sidki}, \citenamefont {Dynes}, \citenamefont {Paraïso},
  \citenamefont {Lucamarini}, \citenamefont {Sharpe}, \citenamefont {Yuan},\
  and\ \citenamefont {Shields}}]{yinguangkuanmen}%
  \BibitemOpen
  \bibfield  {author} {\bibinfo {author} {\bibfnamefont {A.}~\bibnamefont
  {Koehler-Sidki}}, \bibinfo {author} {\bibfnamefont {J.~F.}\ \bibnamefont
  {Dynes}}, \bibinfo {author} {\bibfnamefont {T.~K.}\ \bibnamefont {Paraïso}},
  \bibinfo {author} {\bibfnamefont {M.}~\bibnamefont {Lucamarini}}, \bibinfo
  {author} {\bibfnamefont {A.~W.}\ \bibnamefont {Sharpe}}, \bibinfo {author}
  {\bibfnamefont {Z.~L.}\ \bibnamefont {Yuan}},\ and\ \bibinfo {author}
  {\bibfnamefont {A.~J.}\ \bibnamefont {Shields}},\ }\bibfield  {title}
  {\bibinfo {title} {{Backflashes from fast-gated avalanche photodiodes in
  quantum key distribution}},\ }\href {https://doi.org/10.1063/1.5140548}
  {\bibfield  {journal} {\bibinfo  {journal} {APL}\ }\textbf {\bibinfo {volume}
  {116}},\ \bibinfo {pages} {154001} (\bibinfo {year} {2020})}\BibitemShut
  {NoStop}%
\bibitem [{\citenamefont {Meda}\ \emph {et~al.}(2017)\citenamefont {Meda},
  \citenamefont {Degiovanni}, \citenamefont {Tosi}, \citenamefont {Yuan},
  \citenamefont {Brida},\ and\ \citenamefont {Genovese}}]{2017}%
  \BibitemOpen
  \bibfield  {author} {\bibinfo {author} {\bibfnamefont {A.}~\bibnamefont
  {Meda}}, \bibinfo {author} {\bibfnamefont {I.~P.}\ \bibnamefont
  {Degiovanni}}, \bibinfo {author} {\bibfnamefont {A.}~\bibnamefont {Tosi}},
  \bibinfo {author} {\bibfnamefont {Z.}~\bibnamefont {Yuan}}, \bibinfo {author}
  {\bibfnamefont {G.}~\bibnamefont {Brida}},\ and\ \bibinfo {author}
  {\bibfnamefont {M.}~\bibnamefont {Genovese}},\ }\bibfield  {title} {\bibinfo
  {title} {Quantifying backflash radiation to prevent zero-error attacks in
  quantum key distribution},\ }\href {https://doi.org/10.1038/lsa.2016.261}
  {\bibfield  {journal} {\bibinfo  {journal} {LSA}\ }\textbf {\bibinfo {volume}
  {6}},\ \bibinfo {pages} {e16261} (\bibinfo {year} {2017})}\BibitemShut
  {NoStop}%
\bibitem [{\citenamefont {Pinheiro}\ \emph {et~al.}(2018)\citenamefont
  {Pinheiro}, \citenamefont {Chaiwongkhot}, \citenamefont {Sajeed},
  \citenamefont {Horn}, \citenamefont {Bourgoin}, \citenamefont {Jennewein},
  \citenamefont {L\"{u}tkenhaus},\ and\ \citenamefont {Makarov}}]{2018}%
  \BibitemOpen
  \bibfield  {author} {\bibinfo {author} {\bibfnamefont {P.~V.~P.}\
  \bibnamefont {Pinheiro}}, \bibinfo {author} {\bibfnamefont {P.}~\bibnamefont
  {Chaiwongkhot}}, \bibinfo {author} {\bibfnamefont {S.}~\bibnamefont
  {Sajeed}}, \bibinfo {author} {\bibfnamefont {R.~T.}\ \bibnamefont {Horn}},
  \bibinfo {author} {\bibfnamefont {J.-P.}\ \bibnamefont {Bourgoin}}, \bibinfo
  {author} {\bibfnamefont {T.}~\bibnamefont {Jennewein}}, \bibinfo {author}
  {\bibfnamefont {N.}~\bibnamefont {L\"{u}tkenhaus}},\ and\ \bibinfo {author}
  {\bibfnamefont {V.}~\bibnamefont {Makarov}},\ }\bibfield  {title} {\bibinfo
  {title} {Eavesdropping and countermeasures for backflash side channel in
  quantum cryptography},\ }\href {https://doi.org/10.1364/OE.26.021020}
  {\bibfield  {journal} {\bibinfo  {journal} {Opt. Express}\ }\textbf {\bibinfo
  {volume} {26}},\ \bibinfo {pages} {21020} (\bibinfo {year}
  {2018})}\BibitemShut {NoStop}%
\bibitem [{\citenamefont {Chynoweth}\ and\ \citenamefont {McKay}(1956)}]{1956}%
  \BibitemOpen
  \bibfield  {author} {\bibinfo {author} {\bibfnamefont {A.~G.}\ \bibnamefont
  {Chynoweth}}\ and\ \bibinfo {author} {\bibfnamefont {K.~G.}\ \bibnamefont
  {McKay}},\ }\bibfield  {title} {\bibinfo {title} {Photon emission from
  avalanche breakdown in silicon},\ }\href
  {https://doi.org/10.1103/PhysRev.102.369} {\bibfield  {journal} {\bibinfo
  {journal} {Phys. Rev.}\ }\textbf {\bibinfo {volume} {102}},\ \bibinfo {pages}
  {369} (\bibinfo {year} {1956})}\BibitemShut {NoStop}%
\bibitem [{Note1()}]{Note1}%
  \BibitemOpen
  \bibinfo {note} {In practical operation, achieving the ideal $ER$ is
  challenging due to various factors affecting the measurement. The reduction
  in $ER$ may result from the non-ideal spectral characteristics of the
  backflash photons or their broad spectral distribution, which can produce
  complex responses in wavelength-sensitive components. In addition, imperfect
  coupling between the single-mode fiber and the polarization-maintaining fiber
  flange can further contribute to polarization instability. Through continuous
  optimization and repeated measurements, we acquired a large dataset and
  selected the optimal $ER$ values for each polarization state, as summarized
  in Table~I. Notably, the $ER$ obtained for the $\mathinner {|{V}\rangle }$
  polarization state is slightly lower than that of the other states. This
  difference may arise from polarization dispersion and Fresnel effects, which
  make it more challenging to adjust PC3 to the optimal angle during the
  decoding of the $\mathinner {|{V}\rangle }$ state.}\BibitemShut {Stop}%
\bibitem [{\citenamefont {Wang}(2005)}]{wang2005}%
  \BibitemOpen
  \bibfield  {author} {\bibinfo {author} {\bibfnamefont {X.-B.}\ \bibnamefont
  {Wang}},\ }\bibfield  {title} {\bibinfo {title} {Beating the
  photon-number-splitting attack in practical quantum cryptography},\ }\href
  {https://doi.org/10.1103/PhysRevLett.94.230503} {\bibfield  {journal}
  {\bibinfo  {journal} {Phys. Rev. Lett.}\ }\textbf {\bibinfo {volume} {94}},\
  \bibinfo {pages} {230503} (\bibinfo {year} {2005})}\BibitemShut {NoStop}%
\bibitem [{\citenamefont {Lo}\ \emph {et~al.}(2005{\natexlab{a}})\citenamefont
  {Lo}, \citenamefont {Ma},\ and\ \citenamefont {Chen}}]{lo2005}%
  \BibitemOpen
  \bibfield  {author} {\bibinfo {author} {\bibfnamefont {H.-K.}\ \bibnamefont
  {Lo}}, \bibinfo {author} {\bibfnamefont {X.}~\bibnamefont {Ma}},\ and\
  \bibinfo {author} {\bibfnamefont {K.}~\bibnamefont {Chen}},\ }\bibfield
  {title} {\bibinfo {title} {Decoy state quantum key distribution},\ }\href
  {https://doi.org/10.1103/PhysRevLett.94.230504} {\bibfield  {journal}
  {\bibinfo  {journal} {Phys. Rev. Lett.}\ }\textbf {\bibinfo {volume} {94}},\
  \bibinfo {pages} {230504} (\bibinfo {year} {2005}{\natexlab{a}})}\BibitemShut
  {NoStop}%
\bibitem [{\citenamefont {Ma}\ \emph {et~al.}(2005)\citenamefont {Ma},
  \citenamefont {Qi}, \citenamefont {Zhao},\ and\ \citenamefont
  {Lo}}]{youpian}%
  \BibitemOpen
  \bibfield  {author} {\bibinfo {author} {\bibfnamefont {X.}~\bibnamefont
  {Ma}}, \bibinfo {author} {\bibfnamefont {B.}~\bibnamefont {Qi}}, \bibinfo
  {author} {\bibfnamefont {Y.}~\bibnamefont {Zhao}},\ and\ \bibinfo {author}
  {\bibfnamefont {H.-K.}\ \bibnamefont {Lo}},\ }\bibfield  {title} {\bibinfo
  {title} {Practical decoy state for quantum key distribution},\ }\href
  {https://doi.org/10.1103/PhysRevA.72.012326} {\bibfield  {journal} {\bibinfo
  {journal} {Phys. Rev. A}\ }\textbf {\bibinfo {volume} {72}},\ \bibinfo
  {pages} {012326} (\bibinfo {year} {2005})}\BibitemShut {NoStop}%
\bibitem [{\citenamefont {Wang}\ \emph {et~al.}(2008)\citenamefont {Wang},
  \citenamefont {Peng}, \citenamefont {Zhang}, \citenamefont {Yang},\ and\
  \citenamefont {Pan}}]{youpian2}%
  \BibitemOpen
  \bibfield  {author} {\bibinfo {author} {\bibfnamefont {X.-B.}\ \bibnamefont
  {Wang}}, \bibinfo {author} {\bibfnamefont {C.-Z.}\ \bibnamefont {Peng}},
  \bibinfo {author} {\bibfnamefont {J.}~\bibnamefont {Zhang}}, \bibinfo
  {author} {\bibfnamefont {L.}~\bibnamefont {Yang}},\ and\ \bibinfo {author}
  {\bibfnamefont {J.-W.}\ \bibnamefont {Pan}},\ }\bibfield  {title} {\bibinfo
  {title} {General theory of decoy-state quantum cryptography with source
  errors},\ }\href {https://doi.org/10.1103/PhysRevA.77.042311} {\bibfield
  {journal} {\bibinfo  {journal} {Phys. Rev. A}\ }\textbf {\bibinfo {volume}
  {77}},\ \bibinfo {pages} {042311} (\bibinfo {year} {2008})}\BibitemShut
  {NoStop}%
\bibitem [{\citenamefont {Huttner}\ \emph {et~al.}(1995)\citenamefont
  {Huttner}, \citenamefont {Imoto}, \citenamefont {Gisin},\ and\ \citenamefont
  {Mor}}]{pns}%
  \BibitemOpen
  \bibfield  {author} {\bibinfo {author} {\bibfnamefont {B.}~\bibnamefont
  {Huttner}}, \bibinfo {author} {\bibfnamefont {N.}~\bibnamefont {Imoto}},
  \bibinfo {author} {\bibfnamefont {N.}~\bibnamefont {Gisin}},\ and\ \bibinfo
  {author} {\bibfnamefont {T.}~\bibnamefont {Mor}},\ }\bibfield  {title}
  {\bibinfo {title} {Quantum cryptography with coherent states},\ }\href
  {https://doi.org/10.1103/PhysRevA.51.1863} {\bibfield  {journal} {\bibinfo
  {journal} {Phys. Rev. A}\ }\textbf {\bibinfo {volume} {51}},\ \bibinfo
  {pages} {1863} (\bibinfo {year} {1995})}\BibitemShut {NoStop}%
\bibitem [{\citenamefont {Gottesman}\ \emph {et~al.}(2004)\citenamefont
  {Gottesman}, \citenamefont {Lo}, \citenamefont {Lutkenhaus},\ and\
  \citenamefont {Preskill}}]{2004}%
  \BibitemOpen
  \bibfield  {author} {\bibinfo {author} {\bibfnamefont {D.}~\bibnamefont
  {Gottesman}}, \bibinfo {author} {\bibfnamefont {H.-K.}\ \bibnamefont {Lo}},
  \bibinfo {author} {\bibfnamefont {N.}~\bibnamefont {Lutkenhaus}},\ and\
  \bibinfo {author} {\bibfnamefont {J.}~\bibnamefont {Preskill}},\ }\bibfield
  {title} {\bibinfo {title} {Security of quantum key distribution with
  imperfect devices},\ }in\ \href {https://doi.org/10.1109/ISIT.2004.1365172}
  {\emph {\bibinfo {booktitle} {International Symposium onInformation Theory,
  2004. ISIT 2004. Proceedings.}}}\ (\bibinfo {year} {2004})\ pp.\ \bibinfo
  {pages} {136--}\BibitemShut {NoStop}%
\bibitem [{\citenamefont {Lo}\ \emph {et~al.}(2005{\natexlab{b}})\citenamefont
  {Lo}, \citenamefont {Chau},\ and\ \citenamefont
  {Ardehali}}]{lo2005efficient}%
  \BibitemOpen
  \bibfield  {author} {\bibinfo {author} {\bibfnamefont {H.-K.}\ \bibnamefont
  {Lo}}, \bibinfo {author} {\bibfnamefont {H.~F.}\ \bibnamefont {Chau}},\ and\
  \bibinfo {author} {\bibfnamefont {M.}~\bibnamefont {Ardehali}},\ }\bibfield
  {title} {\bibinfo {title} {Efficient quantum key distribution scheme and a
  proof of its unconditional security},\ }\href
  {https://doi.org/10.1007/s00145-004-0142-y} {\bibfield  {journal} {\bibinfo
  {journal} {J. Cryptol}\ }\textbf {\bibinfo {volume} {18}},\ \bibinfo {pages}
  {133} (\bibinfo {year} {2005}{\natexlab{b}})}\BibitemShut {NoStop}%
\end{thebibliography}%
\end{document}